\documentclass[journal=jctcce,manuscript=article,
layout=twocolumn,email=false]{achemso}
\setkeys{acs}{articletitle=true} 
\usepackage{graphicx}
\usepackage{amsmath}
\usepackage{amsfonts}
\usepackage{amssymb}
\usepackage{hyperref}
\usepackage{dcolumn}
\usepackage{bm}
\usepackage{physics}

\SectionNumbersOn

\DeclareUnicodeCharacter{2212}{-}

\usepackage[utf8]{inputenc}
\usepackage[T1]{fontenc}
\usepackage{mathptmx}

\let\oldmaketitle\maketitle
\let\maketitle\relax

\title{Linear combinations of cluster mean-field states applied to spin systems}

\author{Athanasios Papastathopoulos-Katsaros}
\affiliation{Department of Chemistry, Rice University, Houston, Texas 77005, USA}

\author{Thomas M. Henderson}
\affiliation{Department of Chemistry, Rice University, Houston, Texas 77005, USA}
\alsoaffiliation{Department of Physics and Astronomy, Rice University, Houston, Texas 77005, USA}

\author{Gustavo E. Scuseria}
\affiliation{Department of Chemistry, Rice University, Houston, Texas 77005, USA}
\alsoaffiliation{Department of Physics and Astronomy, Rice University, Houston, Texas 77005, USA}

\begin{document}


\twocolumn[
\begin{@twocolumnfalse}
\oldmaketitle
\begin{abstract}
We present an innovative cluster-based method employing linear combinations of diverse cluster mean-field (cMF) states, and apply it to describe the ground state of strongly-correlated spin systems. In cluster mean-field theory, the ground state wavefunction is expressed as a factorized tensor product of optimized cluster states. While our prior work concentrated on a single cMF tiling, this study removes that constraint by combining different tilings of cMF states. Selection criteria, including translational symmetry and spatial proximity, guide this process. We present benchmark calculations for the one- and two-dimensional $J_1-J_2$ and $XXZ$ Heisenberg models. Our findings highlight two key aspects. First, the method offers a semi-quantitative description of the $0.4 \lessapprox J_2/J_1 \lessapprox 0.6$ regime of the $J_1-J_2$ model - a particularly challenging regime for existing methods. Second, our results demonstrate the capability of our method to provide qualitative descriptions for all the models and regimes considered, establishing it as a valuable reference. However, the inclusion of additional (weak) correlations is necessary for quantitative agreement, and we explore methods to incorporate these extra correlations. 
\end{abstract}

\end{@twocolumnfalse}
]


\section{Introduction}
Hartree-Fock (HF) mean-field theory treats complex systems using a wave function correct for non-interacting electrons. When the correlations between electrons are weak, this approach works well. When those correlations are strong, however, the HF treatment is inadequate. Cluster mean-field (cMF \cite{jimenez-hoyos_cluster-based_2015,papastathopoulos-katsaros_coupled_2022, papastathopoulos-katsaros_symmetry-projected_2023, ghassemi_tabrizi_ground_2023, abraham_cluster_2021, abraham_coupled_2022, abraham_selected_2020, braunscheidel_generalization_2023, braunscheidel_accurate_2024}) theory generalizes this basic idea but provides a more nuanced and flexible framework. It also uses a wave function which is correct for non-interacting constituents, but where in HF these constituents are the individual electrons, cMF uses multi-electronic fragments. Crucially, the wave function within each fragment (which we usually refer to as a ``cluster'', ``tile'', ``plaquette'' or ``covering'') is correlated.  That is, cMF has a wave function which includes correlations within a fragment but not between fragments. This approach seems particularly suited to systems where correlations are in some sense localized, as seen in lattice models with short-range interactions or in systems of well-separated molecules.
\par In recent publications, \cite{papastathopoulos-katsaros_coupled_2022, papastathopoulos-katsaros_symmetry-projected_2023} we have successfully applied cluster-based methodologies to address strongly correlated spin systems, demonstrating their effectiveness in dealing with the challenging $J_1 - J_2$ and $XXZ$ Heisenberg models. These investigations are extensions of our previous work, where we introduced cluster mean-field theory (cMF) for fermionic systems.\cite{jimenez-hoyos_cluster-based_2015} One of the key questions we must ask in cMF is what the individual clusters should be. When the clusters are sufficiently large, the details of the cluster size and shape are not very important. However, because the computational cost of cMF scales exponentially with cluster size, practical considerations require relatively small clusters, and when the clusters are small, cMF is more sensitive to the size and shape of the tiles. We can try to mitigate this dependence to some extent by using inhomogeneous plaquettes,\cite{neuscamman_nonstochastic_2011} but relying on a single underlying tiling scheme with relatively small tiles ultimately limits the accuracy of cMF.
\par Here, we consider a linear combination of cMF states (LC-cMF), each of which uses a different tiling scheme. This leads to a kind of cMF-based nonorthogonal configuration interaction and alleviates the dependence of cMF upon the tiling scheme chosen. Moreover, our results show that this LC-cMF approach provides significant benefits for strongly-correlated systems for which cMF with small tiles is inadequate.
\par The basic idea of combining several tiling schemes is not a new one, but has mostly been restricted to two-site tiles (``dimers'') often associated with valence bond theory.\cite{Weyl1932, pauling_nature_1989} This approach has found extensive applications in the examination of polymers and Heisenberg chains, \cite{klein_many-body_1997, garcia-bach_long-range_2000, garcia-bach_spin-peierls_1997, garcia-bach_cluster-expansion_1996, garcia-bach_valence-bond_1992} and even square-planar lattices. \cite{klein_exact_1990, klein_resonating-valence-bond_1991} Notably, resonating valence bond theory \cite{anderson_resonating_1987} has been put forward as a means to elucidate phenomena such as high-temperature superconductivity, particularly in cuprate compounds, rendering it even more relevant in the context of addressing strong-correlation effects.
\par A parallel endeavor was undertaken by Garcia-Bach and Klein, as evident in works like Refs.~\citenum{klein_exact_1990, klein_resonating-valence-bond_1991}, which are primarily concerned with singlet ground states and are limited to dimer coverings. In contrast, we expand our approach to accommodate diverse tile sizes and systems of varying $S^2$ quantum numbers. This is motivated as we want to explore both $XXZ$ and $J_1 - J_2$ Heisenberg lattices, mainly because the intermediate spin-liquid-like region $J_2/J_1 \approx 0.5$ of the square $J_1 - J_2$ Heisenberg model is very difficult for conventional methods to describe and the $XXZ$ model does not have $S^2$ as a symmetry and comes with a diverse phase spectrum. For a more comprehensive understanding of the interplay between cMF and other advanced methodologies, as well as insights into the advantages of cMF relative to these approaches, readers are directed to Refs.~\citenum{jimenez-hoyos_cluster-based_2015, papastathopoulos-katsaros_coupled_2022,papastathopoulos-katsaros_symmetry-projected_2023} and the associated references therein.
\par Our focus here is on benchmarking cMF-based methods for spin systems because they have lower computational cost than do equivalently-sized fermionic lattices and because exact or nearly exact numerical results are readily available for relatively large spin lattices. It should, however, be noted that cMF has previously been shown to be effective for fermionic systems \cite{jimenez-hoyos_cluster-based_2015, fang_block_2007} as well as the spin systems which we are considering here. To assess the validity of our findings, we compare our results for one-dimensional (1D) systems to those obtained using the density matrix renormalization group (DMRG), \cite{nakatani_matrix_2018} as it gives the exact results in 1D with a manageable, large bond dimension, and to those obtained using exact diagonalization (FCI) for two-dimensional systems.

\begin{figure}
\centering
\includegraphics[scale=0.06]{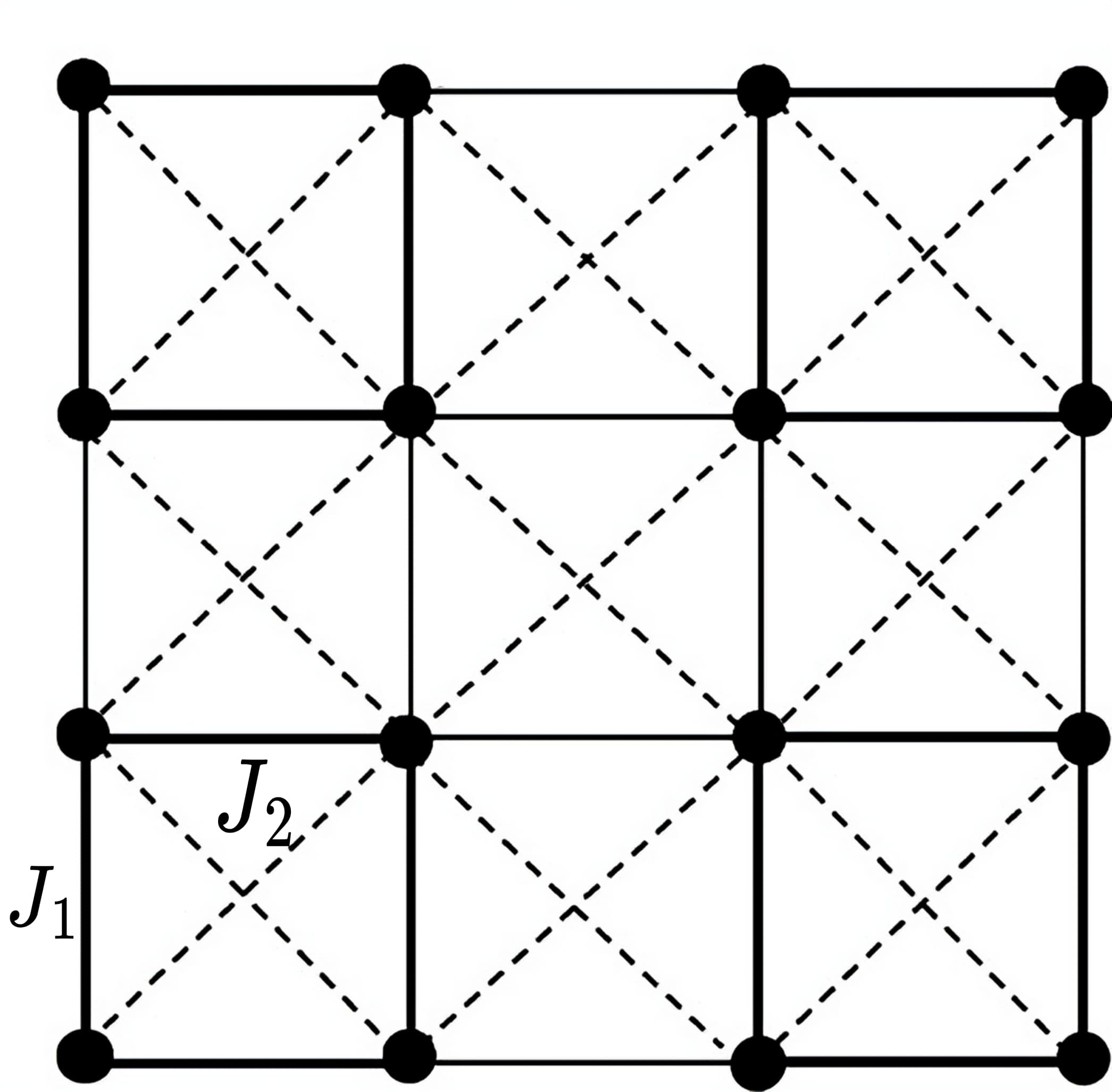}
\caption{Nearest ($J_1$) and next-nearest neighbor ($J_2$) interactions.}
\label{j1j2}
\end{figure}

\section{Background}\label{2.0}
\subsection{Heisenberg model}
Spin lattices, particularly those represented by Heisenberg models, possess considerable chemical significance. An illustration of this is the analysis of iron-sulfur clusters, such as ferredoxins related to nitrogen fixation or photosynthesis, which have been simulated based on the Heisenberg model.\cite{chan} Another instance involves single-chain magnets, like Cobalt(II) Thiocyanate, which have been modeled using the $XXZ$ chain.\cite{rams_singlechain_2020} Furthermore, specific electrides, conjugated hydrocarbons, and superconductors exhibit features reminiscent of Heisenberg exchange interactions.\cite{apps1,apps2,apps3}
\par In this study, we investigate the $XXZ$ and the $J_1$ - $J_2$ Heisenberg models. Both models characterize an assembly of interacting spins on a lattice with a finite size $N$. The $XXZ$ model specifically accounts for interactions solely between nearest neighbors and incorporates anisotropic interactions that violate $S^2$, whereas the $J_1$ - $J_2$ model encompasses both nearest and next-nearest neighbor interactions with isotropic interactions, preserving $S^2$ as a symmetry. In the case of a one-dimensional lattice, both models can be exactly solved using Bethe ansatz.\cite{bethe_zur_1931, yang_ground-state_1966} Although exact solutions are not achievable for the 2D cases, extensive numerical studies have been conducted (refer to, for instance, Refs.~\citenum{dagotto_phase_1989, schulz_finite-size_1992, richter_spin-12_2010, capriotti_spontaneous_2000, mambrini_plaquette_2006, schulz_magnetic_1996, schmalfus_quantum_2006, darradi_ground_2008, bishop_phase_1998, richter_spin-12_2015, bishop_main, jiang_spin_2012, gong_plaquette_2014, murg_exploring_2009, yu_spin-_2012, wang_constructing_2013, capriotti_resonating_2001, sandvik_finite-size_1997} for the $J_1 - J_2$ Hamiltonian and Refs.~\citenum{massaccesi_variational_2021, de_sousa_quantum_2003,cuccoli_two-dimensional_1995, jung_guide_2020, macri_bound_2021, runge_exact-diagonalization_1994,pal_colorful_2021} for the $XXZ$ model).
The Hamiltonian of the $XXZ$ Heisenberg model is
\begin{eqnarray}
H = \sum_{\langle ij \rangle} \bigg [ \frac{1}{2} (S_i^+S_j^- + S_i^-S_j^+) + \Delta S_i^zS_j^z \bigg]
\end{eqnarray}
where $S_i^\pm$ and $S_i^z$ denote the standard spin-$\frac{1}{2}$ operators acting on site $i$, $\langle ij \rangle$ indicates nearest neighbor interactions, and $\Delta$ is the parameter signifying the anisotropy of the model. The exact ground state of the one-dimensional instance of this model exhibits three distinct spin configurations:\cite{massaccesi_variational_2021} For $\Delta \scalebox{0.9}{\ensuremath \gtrsim} 1$, the magnetic correlations manifest as Néel antiferromagnetic; for $\Delta ~ \scalebox{0.9}{\ensuremath \lesssim}~ -1$, they are ferromagnetic; and for $-1 ~ \scalebox{0.9}{\ensuremath\lesssim} ~ \Delta ~ \scalebox{0.9}{\ensuremath\lesssim} ~ 1$, the system resides in the XY phase characterized by gapless excitations and long-range correlations.\cite{schollwock_quantum_2004} Notably, at $\Delta=-1$, the system's ground state is a maximally entangled, extreme antisymmetrized geminal power (AGP) state,\cite{zhiyuan_arxiv, massaccesi_variational_2021} with an energy of $E = -N/4$, where $N$ denotes the number of spins. Lastly, the ferromagnetic phase lends itself to relatively straightforward perturbative treatment, starting from a product state (HF-like) where all spins align in the $z$-direction. Consequently, our focus primarily centers on the more challenging region of $\Delta \geq -1$.
\par
\begin{figure}
\centering
\includegraphics[scale=0.2]{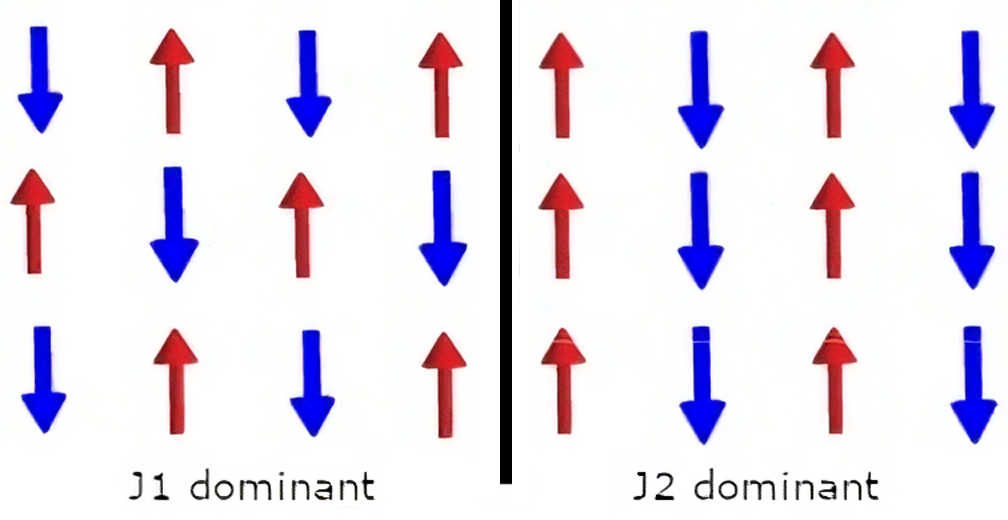}
\caption{Néel (left) and striped (right) antiferromagnetic spin configurations of the square $J_1 - J_2$ Heisenberg model.}
\label{phases}
\end{figure}

 The Hamiltonian of the $J_1$ - $J_2$ Heisenberg model is
\begin{eqnarray}
H = J_1 \sum_{\langle ij \rangle} \vec {S_i} \cdot \vec {S_j} + J_2 \sum_{\langle \langle ij \rangle \rangle} \vec {S_i} \cdot \vec {S_j}
\end{eqnarray}
where $\vec {S_i}$ represents the spin-$\frac{1}{2}$ vector operator at site $i$, while $J_1$ and $J_2$ are the coefficients for nearest-neighbor and next-nearest-neighbor (indicated by $\langle \langle ij \rangle \rangle$) couplings, respectively (refer to Fig.~\ref{j1j2}). For the ensuing discussion, we restrict our focus to the antiferromagnetic (AFM) scenario, specifically when $J_1, J_2 > 0$. It is noteworthy that in the one-dimensional scenario at $J_2/J_1 = 0.5$, known as the Majumdar-Ghosh point, \cite{majumdar_nextnearestneighbor_1969} the exact ground state is a uniform charge-density wave of nearest-neighbor dimers (clusters of two sites). Conversely, the two-dimensional (square) case is more intricate. Over the past two decades, various methods have been employed to study this model extensively. It has been established that in the range $0 ~ \scalebox{0.9}{\ensuremath\lesssim}~ J_2/J_1 \scalebox{0.9}{\ensuremath\lesssim} ~ 0.4$, the ground state exhibits an AFM phase with Néel order, primarily influenced by nearest-neighbor interactions $J_1$. For $J_2/J_1 ~ \scalebox{0.9}{\ensuremath\gtrsim} ~ 0.6$, the ground state demonstrates an AFM phase with striped long-range order, primarily due to the dominance of the next-nearest-neighbor coupling $J_2$ (see Fig.~\ref{phases}). In the range $0.4 ~\scalebox{0.9}{\ensuremath\lesssim} ~ J_2/J_1 \scalebox{0.9}{\ensuremath\lesssim} ~ 0.6$, denoted as the paramagnetic phase, the system encounters frustration as Néel and striped orders compete. The precise nature of this intermediate ground state remains a subject of debate, as do the type of phase transitions and the corresponding transition points (for a more comprehensive discussion, see Refs.~\citenum{schulz_finite-size_1992, gelfand_series_1990, zhitomirsky_valence-bond_1996, takano_nonlinear_2003, isaev_hierarchical_2009, lante_ising_2006, jiang_spin_2012, wang_constructing_2013, hu_direct_2013, li_gapped_2012, nomura_dirac-type_2021, roth_high-accuracy_2022}). The two extreme configurations are depicted schematically in Fig.~\ref{phases}. In finite systems, similar phenomena are observed, although without well-defined transitions. 

\subsection{Cluster Mean-Field}
The approach utilized in this paper for cluster mean-field (cMF) is an extension of our group's prior work outlined in Ref.~\citenum{jimenez-hoyos_cluster-based_2015}. For an in-depth introduction to cluster-based methods, we refer the reader to that reference. Here, we provide a general overview of the framework and its current expansions. 
\par In conventional mean-field theory, the wavefunction is
\begin{eqnarray}
\ket{\Phi_{HF}} = \underset{orbs}{\otimes} \ket{\phi_i}
\end{eqnarray}
where $\phi_i$ represents a state defined in the spin-orbital $i$. Hartree-Fock optimizes the $\phi_i$ to minimize $\bra{\Phi}H\ket{\Phi}$. A parallel approach can be adopted to define the cMF wavefunction:
\begin{eqnarray}
\ket{\Phi_{cMF}} = \underset{clusters}{\otimes} \ket{\phi_i}
\end{eqnarray}
In this case, $\phi_i$ denotes cluster wavefunctions chosen to minimize $\bra{\Phi}H\ket{\Phi}$. It is worth noting that for SU(2) systems when the clusters consist of single sites, cMF simplifies to standard mean-field theory. However, for fermions, the optimization of the the single-site ``tiles'' is necessary for this to hold.\cite{jimenez-hoyos_cluster-based_2015}
\par Cluster mean field provides different results for different clusterization schemes. For example, it is generally more accurate for larger clusters. As a result, specifying the clusterization scheme becomes important. To address this, we introduce a notation system: cMF[$m$] signifies cMF with \textit{m}-site clusters in one dimension, while cMF[$m \times n$] denotes cMF with $m \times n$ clusters in two dimensions, and so forth. When non-rectangular shapes are used, distinct names will be employed to differentiate them. Although cMF allows for clusters of varying sizes, our current focus in this study centers on dimers (clusters of 2) and tetramers (clusters of 4). These choices are favored due to their lower computational demands.
\par Apart from being influenced by the size and configuration of clusters, cMF also relies on the symmetry restrictions applied to these clusters. In this study, we maintain the constraint that each cluster represents an $S_z=0$ eigenstate (where we use the same symbol for the operator and the eigenvalues), consistent with the principles of both restricted (RcMF) and unrestricted cMF (UcMF). Despite our previous findings,\cite{papastathopoulos-katsaros_symmetry-projected_2023} which highlighted the nearly exact nature of generalized cMF (GcMF) around $\Delta=-1$ for the $XXZ$ model, our current emphasis lies in broadening our approach by incorporating additional cluster coverings rather than expanding the Hilbert space for each cluster.

\par For \textit{2}-site clusters which are $S_z = 0$ eigenstates, there are two configurations which we can separate into singlet ($S = 0$) and triplet ($S = 1$) components:
\begin{align}
\label{s0}
\ket{\phi_{ij}}_{S=0} = \frac{1}{\sqrt2}(\ket{\uparrow_i \downarrow_j} ~- ~\ket{\downarrow_i \uparrow_j })
\end{align}

\begin{align}
\label{s1}
\ket{\phi_{ij}}_{S=1} = \frac{1}{\sqrt2}(\ket{\uparrow_i \downarrow_j} ~+ ~\ket{\downarrow_i \uparrow_j })
\end{align}
In RcMF, we choose each cluster wave function to be a singlet. In UcMF, we can choose them to be a linear combination of singlet and triplet, which can also be equivalently defined as
\begin{align}
\label{genucmf}
\ket{\phi_{ij}^{UcMF}} = & ~c_{\uparrow \downarrow}^{ij} ~\ket{\uparrow_i \downarrow_j} ~+ ~c_{\downarrow \uparrow}^{ij} ~\ket{\downarrow_i \uparrow_j }
\end{align} with the requirement of the $c^{ij}$ coefficient optimization. This increases the computational cost, but provides much larger variational freedom. In our work, we will use the former definition for dimers and the more general approach (with coefficient optimization) for tetramers, for which a complete basis cannot be straightforwardly defined as we shall see in section \ref{tet}.

\subsection{Matrix elements and cMF optimization}
The computation of matrix elements in LC-cMF follows a similar procedure to what we have described in our prior publications.\cite{papastathopoulos-katsaros_coupled_2022, papastathopoulos-katsaros_symmetry-projected_2023} Calculating matrix elements involving states with distinct tilings does not significantly increase the computational cost. The overall efficiency of this approach is contingent on the number of coverings included and the optimization strategy for the coefficients (whether by selecting specific singlet and triplet states or by concurrently optimizing the coefficients for each covering).

\subsection{Computational details}
In this study, our computations were conducted using an in-house code, which utilizes the ITensor \cite{itensor} library to construct the cMF states, simplifying the computation of necessary matrix elements. In the tetramer scenario, we initiated the cluster states with random configurations and employed the conjugate gradient \cite{atkinson_introduction_1989} algorithm with numerical gradients to optimize the energy. For conducting the NOCI, we utilized the generalized eigenvalue solver from the GNU Scientific Library (GSL).
\par As previously mentioned, we have employed dimers (clusters of 2 sites) and tetramers (clusters of 4 sites) in our research. These compact clusters are computationally efficient in the cMF framework and conveniently form the fundamental components of our lattices. It is worth noting that the specific shapes of these clusters might not always yield the lowest LC-cMF energy, primarily due to finite-size effects. Nevertheless, when the clusters reach a sufficient size, their shape becomes less significant.
\par All calculations utilized an equal count of up and down spins within each cluster. This practice was employed to ensure that the clusters represented $S_z=0$ eigenfunctions. 

\section{LC-cMF}\label{3.0}
\subsection{A complete basis of dimer coverings}
A ``covering'' or ``tiling'' is defined as a product of cluster states with non-overlapping indices. It has been shown \cite{garcia-bach_spin-peierls_1997} that a linear combination of a combinatorial number of linearly independent dimer coverings of valence bonds (RcMF[2]) with optimized coefficients can give the exact singlet ground state of Heisenberg lattices. Mathematically,
\begin{align}
\ket{\Psi_{FCI}} = \sum_a^n c_a\ket{C_a}
\end{align}
where $C_a$ is an RcMF[2] state for the $a$ dimer covering. For the 1D case, $n$ is given by
\begin{align}
n = \frac{N!}{(N/2+1)!(N/2)!}
\end{align}
where $N$ is the number of sites. Finding these dimer coverings is non-trivial and constitutes a problem in graph theory and statistical physics.\cite{kasteleyn_statistics_1961} Below we will show the algorithm for the 1D case, and for the interested reader, the 2D case is extensively covered in Refs.~\citenum{klein_exact_1990,klein_resonating-valence-bond_1991}.

\subsection*{Finding the dimer coverings for the 1D case}
\begin{figure}
\centering
\includegraphics[scale=0.3]{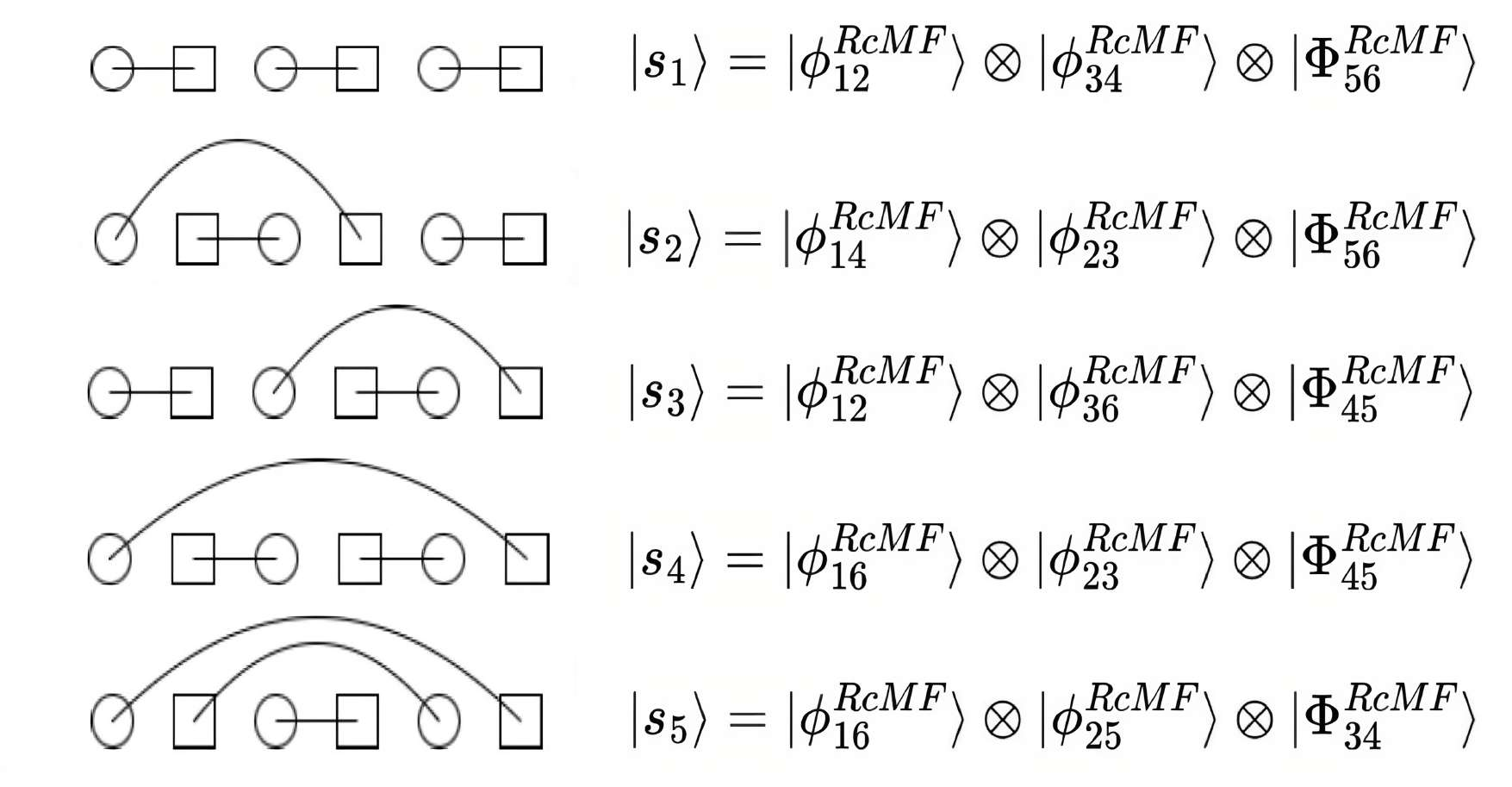}
\caption{All 5 dimer coverings for the 6-site Heisenberg chain. Figure adapted from Ref.~\citenum{garcia-bach_spin-peierls_1997}.}
\label{algorithm}
\end{figure}

There are many ways of finding the combinatorial number of linearly independent dimer coverings, but in this work, we focus on the following algorithm, depicted in Fig.~\ref{algorithm} with an example for 6 sites (5 dimer coverings):
\begin{enumerate}
    \item We start by forming a bipartite lattice using circles and squares.
    \item We match circles to squares, but with one caveat: the lines cannot cross. 
\end{enumerate}
 If we include all coverings and perform a non-orthogonal configuration interaction (NOCI) calculation, we get the exact ground state energy.

\subsection{Truncating the covering basis}
The full basis of all permitted coverings is identical in size to the $S=0$ basis of the Hamiltonian and therefore the NOCI has the same cost as the FCI in the symmetry-adapted sector. In practical calculations for large lattices, we can only use a fraction of the available covering patterns, and the question then becomes how to best select a truncated basis of coverings. In doing so, we will focus on the case of dimer coverings, which are somewhat simpler, but the ideas we discuss below can be extended to other tile sizes. Regardless, the task is rather challenging because the optimal basis of tilings would be expected to vary depending on the specifics of the Hamiltonian.
\par To address this issue, we begin by noting that the Hamiltonian only directly couples nearest-neighbor and next-nearest-neighbor sites. We therefore expect the most significant correlations to be relatively short range, suggesting that the most important tilings should be fairly localized. For this reason, we opted for a strategy involving the selection of tilings characterized by relatively short ``bond lengths'', where ``bond length'' represents the summation of distances between sites within each dimer for all the dimers in a given covering. Thus, a nearest-neighbor dimer contributes a bond length of 1, a third-nearest-neighbor dimer contributes a bond length of 3, and so forth (recall that there are no next-nearest-neighbor dimers by virtue of the bipartition of the lattice).
\par Once we have established the various tilings, we can order them by their bond lengths. Table \ref{bls} shows the results of this process for the dimer coverings in Fig.~\ref{algorithm}.  We can then define a ``bond level'' which denotes the position of the bond length within this ordering. In the context of Fig.~\ref{algorithm}, bond level 1 refers to tiling $\ket{s_1}$, bond level 2 refers to tilings $\ket{s_2}$ and $\ket{s_3}$, and so on. We can now define a truncation wherein we retain all tilings within a certain bond level. Increasing the bond level increases the number of tilings we include, and we always include the most local tilings first. We symbolize this truncation scheme as $d$LC-cMF[$n$], where $d$ refers to the bond level and $n$ to the cluster size.  For example, 3LC-cMF[2] refers to a computation utilizing all dimer coverings ($n=2$) with a bond level of 3 or less ($d = 3$).
\par For 1D chains with open boundary conditions (OBC), the number of dimer coverings with bond level of 1 scales as $\mathcal{O}(N)$ while for bond level of 2 it scales as $\mathcal{O}(N^2)$. For lattices with periodic boundary conditions (PBC) and 2D lattices, computing the bond length needs to take the periodicity into account. For example, in PBC, the first and the last site have only 1 site distance, the second to last with the first only 2, etc., whereas, in 2D lattices, the length can be computed by using the smallest bond length while moving along each axis of the lattice. Again, the bond level is defined similarly, regardless of the actual value of the bond lengths. 
\par Finally, we should note where cMF is extensive (the energy scales linearly with system size), LC-cMF as we have employed it here is not.  While extensivity is an important consideration for calculations which seek to reach the thermodynamic limit, it is less important for the relatively small systems we have considered in this work. Generalizations of the schemes we have introduced here to reduce or eliminate extensivity errors are of course highly desirable and are under investigation.
\vspace{-\parskip}
\subsection*{LC-cMF[2] results for 1D}
\par
\begin{figure}
\centering
\includegraphics[scale=1.0]{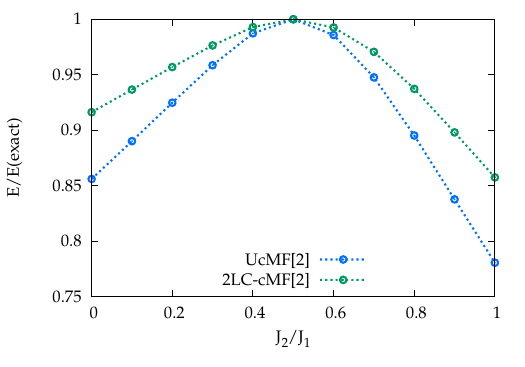}
\caption{Energy comparison of UcMF[2] (UcMF with 2-site tiles) and 2LC-cMF[2] (a linear combination of RcMF with 2-site tiles in 8 different tiling schemes). The results are for the 16-site $J_1 - J_2$ Heisenberg chain.}
\label{1d_j1j2_lccmf}
\end{figure}

 To demonstrate the effectiveness of the dimer covering basis, we begin our analysis by computing the ground state energy of a 16-site $J_1-J_2$ Heisenberg chain with OBC. The results are presented in Fig.~\ref{1d_j1j2_lccmf}. We remind the reader that 2-LC means we include the nearest neighbor tiling and the $N$ tilings with equivalent next shortest bond lengths overall. As mentioned in Sec.~\ref{2.0}, the exact ground state at the Majumdar-Ghosh point ($J_2/J_1 = 0.5$) is a nearest-neighbor dimer covering, hence our method also ought to be exact. However, it is also evident that even with a minimal bond level, our results exhibit qualitatively good agreement with the exact ground state energy, particularly in the vicinity of $J_2/J_1=0.5$. Moreover, the improvement in comparison to a UcMF[2] calculation, where the coefficients are optimized, is significant, underscoring the considerable potential of LC-cMF, which comes without the need of any additional ``correlation'' methods; the correlations are described by a simple linear combination of tilings. Finally, it is worth noting that while UcMF[2] has broken $S^2$ symmetry, 2LC-cMF[2] is an $S^2$ eigenstate. Consequently, this energy improvement comes with the correct quantum numbers, in contrast to UcMF[2].

\begin{table}[h!]
\centering
\begin{tabular}{|c|c|}
\hline
Dimer covering & Bond length \\ \hline
$\ket{s_1}$             & 3           \\ \hline
$\ket{s_2}$             & 5           \\ \hline
$\ket{s_3}$            & 5           \\ \hline
$\ket{s_4}$             & 7           \\ \hline
$\ket{s_5}$             & 9           \\ \hline
\end{tabular}
\caption{Bond lengths of all dimer coverings of Fig.~\ref{algorithm}.}
\label{bls}
\end{table}

\subsection*{LC-cMF[2] results for 2D}\label{4.2}
\begin{figure}
\centering
\includegraphics[scale=1.0]{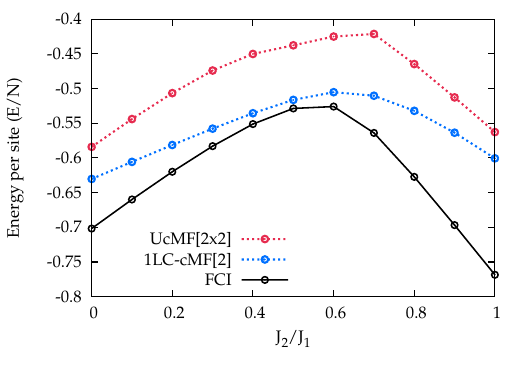}
\caption{Energy per site obtained in UcMF[$2 \times 2$] (UcMF with $2 \times 2$ tiles), 2LC-cMF[2] (a linear combination of RcMF with 2-site tiles in 115 different tiling schemes) and FCI for the $4 \times 4$ $J_1 - J_2$ Heisenberg lattice.}
\label{2d_j1j2_lccmf}
\end{figure}

To illustrate that dimer coverings can also be effectively applied to 2D systems, we take the example of the $4 \times 4$ $J_1 - J_2$ Heisenberg square with PBC, and our findings are presented in Fig.~\ref{2d_j1j2_lccmf}. An intriguing observation is that, once again, we achieve remarkable accuracy for such a simple wave function, especially when compared to UcMF[$2 \times 2$], which has the flexibility of coefficient optimization. This underscores that, with relatively small clusters, LC-cMF can outperform the results obtained with larger clusters. In particular, LC-cMF[2] excels in the non-magnetic $J_2/J_1 \approx 0.5$ region, which is the region that is most difficult to describe. This is probably a spin-liquid-like feature that this regime has, which can be captured by this ansatz.

\section{Correlation and non-singlet cases}\label{4.0}
The methodology presented in the previous section limits us to spin systems which have $S^2$ as a symmetry, because each tile, and therefore the system as a whole, is an $S^2$ eigenstate with $S^2 = 0$. In this section, we extend this methodology to systems which do not have this symmetry, and simultaneously incorporate correlations within the cMF scheme, similarly to what is done for cluster-CI (cCI).\cite{abraham_selected_2020, braunscheidel_generalization_2023, braunscheidel_accurate_2024}

\subsection{Dimer case}
For non-singlet cases, we could consider a simple linear combination of UcMF states (as defined in Eqn.~\ref{genucmf}). This would require us to optimize the coefficients $c^{ij}_{\uparrow \downarrow}$ and $c^{ij}_{\downarrow \uparrow}$, in each tile, for each tiling scheme in the linear combination.  To simplify the calculation and reduce the cost, we try a different approach. For a given tiling scheme, we include the RcMF state and a handful of excited states in which one or more clusters have been excited from the singlet configuration (Eqn.~\ref{s0}) to the triplet (Eqn.~\ref{s1}). For Hamiltonians which do not have $S^2$ symmetry, such as the XXZ model, this procedure is needed so that the wave function does not have a symmetry it should not possess. For a given tiling with $N$ \textit{2}-site tiles, this gives us $2^N$ states. In practice we will generally wish to limit the number of such excitations. In addition to the RcMF state, we will include all cases in which one or two tiles are in the triplet configuration, which we refer to as ``single'' and ``double'' excitations, respectively. This follows the usual terminology for cluster-based configuration interaction where a single excitation has one cluster not in its ground state, a double excitation has two clusters not in the ground state, and so on. 
\par We may choose to include higher excitations, but if we do, we restrict the excited clusters to all be adjacent to one another, similarly to Ref.~\citenum{bishop_phase_1998}. We symbolize this truncation scheme as dLC-cMF[n]-exc, where $d$ refers to the bond level and $n$ refers to the cluster size as noted earlier, and $exc$ refers to the triplet states. For example, 3LC-cMF[2]-sd refers to a computation utilizing all dimer coverings with bond level of 3, with all the one-at-a-time and two-at-a-time triplet cases. Lastly, we note that for Hamiltonians with $S^2$ symmetry, e.g. the $J_1 - J_2$ Heisenberg model, we do not require single excitations as they inevitably change the $S^2$ quantum number. 
\par
\begin{figure}
\centering
\includegraphics[scale=1.0]{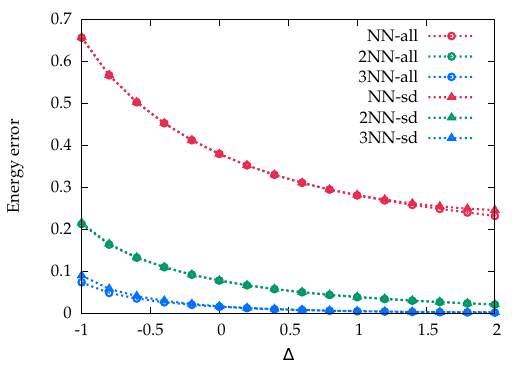}
\caption{Comparison of energy errors of different bond levels and triplet excitations for LC-cMF[2] (linear combinations of RcMF with 2-site tiles in different tiling schemes). All calculations are performed on the 16-site $XXZ$ Heisenberg chain.}
\label{lccmf_cisd}
\end{figure}

\begin{figure}
\centering
\includegraphics[scale=1.0]{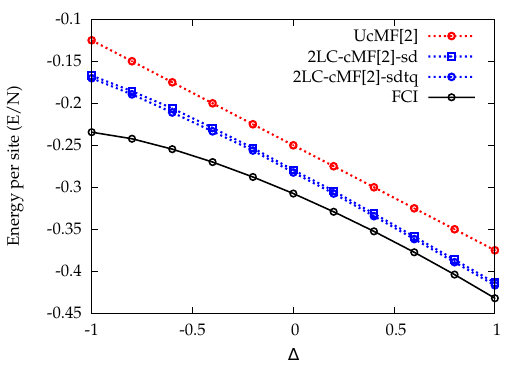}
\caption{Energy per site obtained in UcMF[2], 2LC-cMF[2]-sd (a linear combination of RcMF with 2-site tiles in 18 different tiling schemes and up to two-at-a-time triplets), 2LC-cMF[2]-sdtq (a linear combination of RcMF with 2-site tiles in 18 different tiling schemes and up to four-at-a-time triplets) and FCI for the 36-site $XXZ$ Heisenberg chain.}
\label{1d_xxz_lccmf}
\end{figure}

 To demonstrate the efficiency of this correlation and truncation scheme, we present results for 1D XXZ Heisenberg systems with OBC in Figs.\ref{lccmf_cisd} and \ref{1d_xxz_lccmf}. In Fig.\ref{lccmf_cisd}, our results for the 16-site XXZ Heiseinberg chain with OBC are shown. As expected, including only single and double excitations yields results virtually indistinguishable from including all excitations. This result is not exact because each tile is still forced to have $S_z = 0$, and the case where a given tile has $S_z = \pm 1$ has been totally excluded. These extra tilings would be included in a GcMF\cite{papastathopoulos-katsaros_symmetry-projected_2023} version of the theory, which we elect not to pursue in the present work. Nevertheless, this outcome is expected since the crucial part of the wavefunction arises from the various tiling schemes, with triplet excitations introducing secondary effects. Additionally, given the relatively small system size, we do not exclude many excitations. As a result, for much larger systems, we opt to include up to quadruple excitations, but triple and quadruple excitations are allowed only for contiguous dimers, as we have already noted. Our results for the 36-site $XXZ$ chain with OBC are summarized in Fig.\ref{1d_xxz_lccmf}. Here, we can achieve an accurate approximation to the exact solution, especially for values of $\Delta > 0$, despite the small bond length. The discrepancies around $\Delta=-1$ are not a significant concern, as our previous work \cite{papastathopoulos-katsaros_symmetry-projected_2023} has demonstrated how to obtain highly accurate energy estimates in that regime. It is possible to imagine a linear combination of UcMF or even GcMF states, which would come with greater variatonal flexibility and should yield superior answers albeit with a significantly greater computational cost. Alternatively, since a single GcMF suffices for the region for $\Delta < -1$, we could also just include this one GcMF state in our linear combination.

\subsection{Tetramer case}\label{tet}
\begin{figure}
\centering
\includegraphics[scale=0.2]{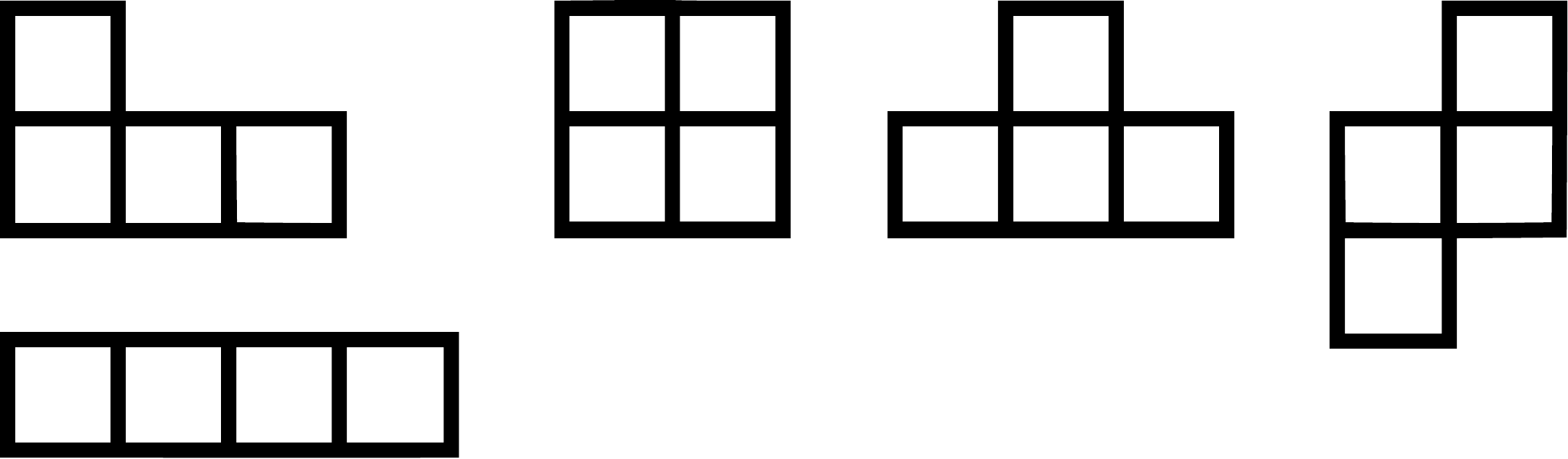}
\caption{The 5 tetramer coverings used for 2D square lattices.}.
\label{tetcov}
\end{figure}

In the present section, we delve into the exploration of linear combinations involving larger clusters, specifically focusing on tetramers due to their viability for the systems we are examining. To the best of our knowledge, there is currently no readily available basis for tetramer coverings in closed form. However, this is not a significant constraint since our methods are designed with a truncation approach in mind. Previous research\cite{jimenez-hoyos_cluster-based_2015} has demonstrated that compact shapes generally yield more accurate estimates of the ground state energy. For this reason, we limit our consideration to 5 tetramer tiles, akin to those depicted in Fig.~\ref{tetcov}, which constitute the fundamental Tetris\cite{tetris} pieces. Each of these 5 tetramer configurations employs a single tile type, albeit with variations in inversion, rotation, and reflection. While it is possible to conceive shapes that, when considering the spin configurations of the exact ground state for each Hamiltonian, might yield lower energies, our objective is to establish a method that operates in a black-box manner, and tiles with intricate shapes can pose challenges.

\par An unrestricted tetramer contains ${4 \choose 2} = 6$ $S_z = 0$ states and can be written as
\begin{align}
\ket{\phi_{ijkl}^{UcMF}} =& ~c_{\uparrow \uparrow \downarrow \downarrow}^{ijkl} ~\ket{\uparrow \uparrow \downarrow \downarrow} ~+ ~c_{\downarrow \downarrow \uparrow \uparrow}^{ijkl} ~\ket{\downarrow \downarrow \uparrow \uparrow} \\ \nonumber +& ~c_{\uparrow \downarrow \uparrow \downarrow}^{ijkl} ~\ket{\uparrow \downarrow \uparrow \downarrow}  ~+ ~c_{\downarrow \uparrow \downarrow \uparrow}^{ijkl} ~\ket{\downarrow \uparrow \downarrow \uparrow}  \\ \nonumber +& ~c_{\uparrow \downarrow \downarrow \uparrow}^{ijkl} ~\ket{\uparrow \downarrow \downarrow \uparrow}  ~+ ~c_{\downarrow \uparrow \uparrow \downarrow}^{ijkl} ~\ket{\downarrow \uparrow \uparrow \downarrow}
\end{align}
where dimers have only one singlet state, in tetramers there are two (as well as three triplets and a quintet). This means that even for a restricted LC-cMF we would have to optimize the linear combination of the two singlet states used in each tile. We therefore decided to optimize all six of the coefficients $c^{ijkl}$ for every tile in every tetramer covering simultaneously. This increases the variational flexibility in our LC-cMF significantly, although with concomitant increase in computational cost. We note, though, that the excitation-based scheme we introduced in the previous section could be employed for tetramers as well, except that where a dimer has only one excited state, a tetramer has five.

\subsection*{LC-cMF[4] results for 2D}\label{4.3}
\begin{figure}
\centering
\includegraphics[scale=1.0]{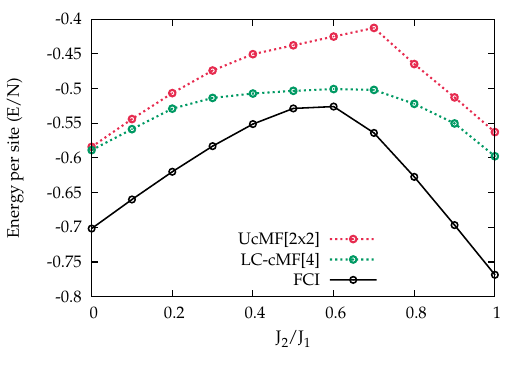}
\caption{Energy per site obtained in UcMF[$2 \times 2$] (UcMF with $2 \times 2$ tiles), LC-cMF[4] (a linear combination of UcMF with 4-site tiles in the 5 Tetris tiling schemes) and FCI for the $4 \times 4$ $J_1 - J_2$ Heisenberg lattice.}
\label{2d_j1j2_lccmf_tetramers}
\end{figure}

 Here we center our results on the $4 \times 4$ $J_1 - J_2$ Heisenberg square with PBC and the outcomes are depicted in Fig.\ref{2d_j1j2_lccmf_tetramers}. Notably, even with just 5 tilings, we achieve reasonable precision once again. Particularly in the non-magnetic $J_2/J_1 \approx 0.5$ region, our results exhibit good accuracy, suggesting the spin liquid-like character of this regime, is encapsulated within our model.

\section{Discussion}\label{5.0}
In section \ref{2.0}, we explored the cluster mean-field method for addressing strongly-correlated spin systems. Building upon this foundation, in section \ref{3.0} we introduced the linear combinations of cluster mean-field formalism for singlet ground states. In section \ref{4.0}, we extended this approach to arbitrary $S^2$, which further allowed us to incorporate additional correlations. Depending on the Hamiltonian, these extensions can be applied as variational approaches for the ground state wavefunction, providing relatively accurate estimates for the energy. Our findings clearly illustrate that LC-cMF methods can qualitatively capture the essential aspects of the ground state physics in benchmark models and can serve as valuable references for more advanced correlated methods, with mild computational cost in an efficient code. In addition, they offer compelling evidence that our method can be employed for a semi-quantitative description of the $0.4 \lessapprox J_2/J_1 \lessapprox 0.6$ regime of the 2D $J_1 - J_2$ model, which is a particularly demanding regime for established methodologies.
\par In our earlier work, we discussed the correlation of GcMF, as detailed in Ref.~\citenum{papastathopoulos-katsaros_symmetry-projected_2023}, with the use of Jastrow-like operators. These operators have been applied previously in the context of the antisymmetrized geminal power (AGP) wavefunction, as highlighted in references such as \citenum{khamoshi_exploring_2021} and \citenum{henderson_correlating_2020}, showcasing promising outcomes. The inclusion of such operators has the potential to enhance the accuracy of our cluster-based wavefunctions and expand their applicability to more intricate systems. Lastly, we can use a similar approach to the few determinant approximation (FED), \cite{rodriguez-guzman_variational_2014,bytautas_potential_2014, rodriguez-guzman_multireference_2013, jimenez-hoyos_multi-component_2013, rodriguez-guzman_multireference_2014, rodriguez-guzman_symmetry-projected_2012} but instead of determinants we can utilize excited cluster product states and optimize them in a NOCI manner, similarly to the present work.
\par While our findings have been specifically focused on spin lattices, they strongly indicate the promise of cMF for these complex model Hamiltonians. It is worth underscoring that the fundamental methodologies we have detailed in this study can readily extend to broader chemical systems. For instance, one could envision utilizing distinct clustering schemes to represent various functional groups within a sizable molecule, where each group of atoms may fulfill unique roles in a specific scheme. The outcomes we have presented here highlight the potential value of these techniques across a wide range of applications.

\section*{Author Information}
\subsection*{Corresponding Author}
\textbf{Gustavo E. Scuseria} − Rice University, Houston, Texas 77005, USA; Email: guscus@rice.edu

\begin{acknowledgement}
This work was supported by the U.S. Department of Energy, Office of Basic Energy Sciences, Computational and Theoretical Chemistry Program under Award DE-SC0001474. G.E.S. acknowledges support as a Welch Foundation Chair (Grant No. C-0036).
\end{acknowledgement}



\bibliography{cc}


\end{document}